# REFORGE: A METHOD FOR BENCHMARKING LLMS' REVERSE ENGINEERING CAPABILITIES IN DECOMPILED BINARY FUNCTION NAMING

Nicolas Koller
*Systrion GmbH*
*Flughafenstrasse 52, D-22335 Hamburg, Germany*

Andreas U. Schmidt
*Wilhelm Büchner Hochschule*
*Department of Computer Science,*
*Hilpertstrasse 31, D-64295 Darmstadt, Germany*

**ABSTRACT**

Large language models (LLMs) are increasingly applied to reverse-engineering tasks, and recent threat-intelligence reporting shows them operating inside live offensive-security workflows. Claims about their capability, however, outpace our ability to measure it. Existing benchmarks for LLM-assisted binary analysis treat the construction of function-level ground truth as a solved pre-processing step and report accuracy without disclosing how many functions were reliably evaluable. We argue that the principal obstacle to fair evaluation is not model capability but the reliability of binary-to-source alignment under compiler optimization. This paper presents Reforge, a provenance-tracked pipeline that constructs function-level ground truth from C source through compilation, DWARF and syntactic extraction, alignment, and decompilation, and that operationalizes alignment uncertainty as an eight-gate confidence funnel with three-tier stratification. On a controlled micro-benchmark, high-confidence yield falls from 87.2% to 65.9% across optimization levels, and unpaired comparisons overstate optimization-induced performance decay through survivorship bias. A proof-of-concept evaluation of seven contemporary LLMs on function naming demonstrates the validity of the concept and generally motivates an uncertainty-aware benchmarking practice.

## 1. INTRODUCTION

Reverse engineering (RE) is the practice of recovering the structure, behaviour, and intent of a program from its compiled form. It underpins vulnerability research, malware analysis, interoperability work, and software forensics, and it is among the most labour-intensive activities in computer security, because compilation discards the names, types, and structure that make source code intelligible. Capable LLMs trained on code have raised the prospect for automation, and a growing body of research applies LLMs to decompiled-code understanding, identifier recovery, and decompilation repair.

The stakes of understanding what LLMs can and cannot do have risen sharply. In November 2025, Anthropic reported what it described as the first documented cyber-espionage campaign executed largely autonomously by an AI system, in which an agent was manipulated into performing reconnaissance, vulnerability discovery, exploitation, and post-exploitation across roughly thirty target organizations with limited human intervention (Anthropic, 2025). Notably, this report observed that model's hallucinations and overstated findings remained a practical limiting factor. LLMs are therefore no longer a speculative tool in the offensive and defensive security loop but an operational one whose reliability is uncertain, which makes the rigorous measurement of LLM capability on core RE subtasks a question of immediate practical importance.

Yet measurement is where the field is weakest. Evaluating an LLM on a task such as recovering a function name from its decompiled, stripped form requires ground truth: a reliable mapping from each binary function back to the source-level identifier it originated from. Constructing that mapping is hard. Compilation is a lossy, many-to-one transformation; optimization inlines, merges, reorders, and eliminates functions, so that the population of source functions and the population of binary functions diverge as optimization increases, and debug-derived mappings fragment under these transformations (Dramko et al., 2024; Tan et al., 2025). Most existing evaluations treat this alignment problem as a pre-processing detail and report a single accuracy number per optimization level without disclosing how many functions were evaluable, how the mapping was validated, or how much of the apparent effect of optimization is an artifact of which functions survived to be measured.

This paper takes the position that the primary bottleneck in benchmarking LLM-assisted reverse engineering is not model capability but the construction and auditing of the evaluation base data itself. We develop a method that makes alignment quality a reportable component of the benchmark rather than an implicit assumption. Concretely, we contribute: (1) Reforge, a provenance-tracked pipeline whose ground-truth construction (build, oracle extraction, and alignment) is deterministic and whose decompilation and model-query stages are treated as documented uncertainty sources, recording every alignment decision so that yield and attrition are auditable; (2) an eight-gate confidence funnel that operationalizes alignment uncertainty as an explicit sequence of quality checks, stratifying functions into three confidence tiers for uncertainty-aware reporting; and (3) evidence, obtained through the method, that unpaired optimization-level comparisons overstate performance decay through survivorship bias, with a concrete remedy in the form of paired analysis over a stable, source-anchored function key. As a fourth, supporting contribution, we report a proof-of-concept evaluation of seven contemporary LLMs on function naming that demonstrates the substrate end to end; its empirical scope is deliberately narrow, and we treat the resulting performance figures as illustrative rather than as general capability estimates. The proof-of-concept evaluation is structured around three research questions. RQ1: how much reliable ground truth survives compiler optimization, and does the resulting attrition bias optimization-level comparisons? RQ2: how accurately do contemporary LLMs recover function names from decompiled, stripped code on this substrate? RQ3: how much additional benefit does a small ranked shortlist of candidate names offer over a single top prediction in an analyst-assistance setting? Sections 4.2 to 4.4 address these questions in turn.

The remainder of the paper reviews the relevant background and the transparency gap in current benchmarks (Section 2); describes the Reforge method, covering the pipeline, dataset, confidence funnel, stable-key paired-analysis design, and metrics (Section 3); reports the proof-of-concept evaluation (Section 4); and discusses what generalizes before concluding with directions for further work (Sections 5 and 6).

## 2. BACKGROUND AND RELATED WORK

A C compiler transforms source into an executable through pre-processing, compilation to assembly, assembly to object code, and linking. Each stage discards information that is irrelevant to execution but essential to human comprehension. High-level names, comments, and types are not represented in machine code; with optimization enabled, the compiler additionally inlines callees into callers, merges identical code, hoists and reorders computation, and eliminates dead or unreachable functions. The standard Linux executable format, ELF, can optionally carry DWARF debugger information that records the source name, declaration location, and address range of each function, but production binaries are typically stripped of all such symbols. The reverse path, decompilation, reconstructs approximate C-like pseudocode from machine instructions, but it cannot restore the discarded identifiers and only partially recovers structure. The practical consequence is an information-loss chain in which the link between a binary function and its originating source declaration becomes progressively weaker and, at higher optimization levels, ambiguous or absent. Because compilation is many-to-one, a unique correct source for a given binary is not guaranteed (Liu et al., 2024).

Work applying neural models to reverse engineering can be grouped by the information the model receives and the task it performs. A first line trains task-specific models to recover identifiers from decompiled code: DIRTY and DIRECT framed variable and type recovery as sequence transduction over decompiler output (Chen et al., 2022; Nitin et al., 2021), and later work extended naming to assembly and to architecture-agnostic representations, or reframed it as contrastive captioning of binary functions (Benoit et al., 2025). A second line evaluates general-purpose LLMs without fine-tuning; benchmarks of GPT-class models on binary code summarization find useful high-level descriptions but weak fine-grained identifier recovery (Jin et al., 2023), and domain-adapted LLMs have been applied specifically to function-name inference in stripped binaries (Jiang et al., 2025). A third line uses LLMs to repair or re-execute decompiler output, targeting recompilable code rather than analyst-facing names (Wong et al., 2025). Across these lines, a recurring framing distinguishes autonomous recovery, which aims to replace the analyst, from analyst assistance, which aims to inform human judgement under uncertainty (Leger et al., 2021); the latter motivates evaluation designs that reward producing (a ranked set of) plausible candidates rather than only producing a single exact answer.

As a common thread, these evaluations report model accuracy without reporting the reliability of the ground truth against which accuracy is measured. Function-level benchmarks are typically assembled by matching binary functions to source via debug information or symbol tables, then filtering noisy pairs (Tan et al., 2025). The matching and filtering steps are consequential (they determine which functions are evaluable, and therefore which population the reported accuracy describes), yet they are rarely audited or reported. Recent benchmarking work has begun to consider this problem in adjacent tasks: type-inference results in decompilers depend heavily on unstated assumptions about type compatibility (Soni et al., 2025), and broader binary-analysis benchmarks for LLMs are emerging (Shang et al., 2025; Ghimire et al., 2025). The Exploring Explicit Uncertainty for Binary Analysis (EUBA) framework articulates the underlying principle: uncertainty in binary analysis is multi-source and compounds across tools and stages, so conflating tool-introduced error with model error invalidates conclusions (Leger et al., 2021). The method described in the present paper is designed around this principle, making alignment yield, per-stage attrition, and per-function confidence explicit, auditable outputs of the benchmark rather than assumptions buried in pre-processing.

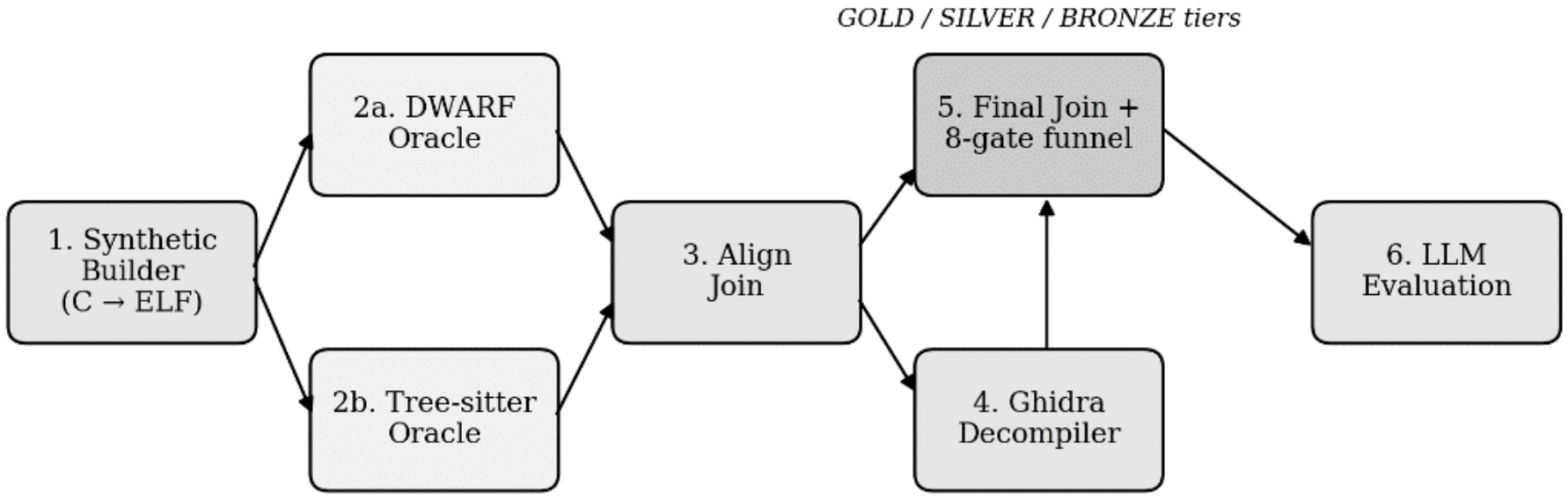

Figure 1. The six-stage Reforge pipeline. Two independent oracles produce ground truth that is aligned, joined to decompiler output, gated, and submitted to LLMs for scoring.

# 3. THE REFORGE METHOD

## 3.1 Design Principles and Pipeline Overview

**Reforge** is a modular pipeline that transforms synthetic C programs through six versioned processing stages, illustrated in Figure 1. Its guiding principle is not that the whole pipeline is deterministic, but that each stage is either made reproducible or has its residual uncertainty documented, following the observation that uncertainty in binary analysis accumulates across a cascade of tools (Leger et al., 2021). The ground-truth core, comprising the builder, the two oracles, and the joins, is deterministic. That is, given identical inputs and a pinned toolchain, each of these stages produces byte-identical outputs (sorted keys, stably ordered records, no random number generation), verified by repeated execution and hash comparison. The two remaining stages carry no such guarantee. The Ghidra decompiler is only conditionally deterministic: although its analysis pipeline is intended to be reproducible for a fixed binary, configuration, and version, its decompilation output is not guaranteed reproducible across runs in every processor module[1] (we do not audit its internals). The LLM

[1]Non-determinism has been documented for at least one Ghidra processor module: repeated C exports of the same MIPS binary differed across runs, whereas x86, x86-64, and ARM exports were reported deterministic in the same report (National Security Agency, 2022). The binaries evaluated here are x86-64, but the pipeline does not depend on this and records decompiler output as a potential uncertainty source regardless. A direct check confirms this for the present corpus: decompiling all 60 stripped binaries twice in independent headless runs under a pinned Ghidra 12.0.3

stage is stochastic by construction. Reforge therefore treats decompilation and model querying as uncertainty sources to be measured and recorded rather than eliminated. No function is silently discarded at any stage: functions that fail a quality check are retained with the reason for failure recorded, so that attrition can be quantified rather than assumed away.

The six stages are as follows. Stage 1, the synthetic builder, compiles each program across a controlled build matrix and emits a provenance-anchored build receipt. Stage 2 runs two independent oracles: a DWARF oracle (2a) that extracts function names, address ranges, and source declaration locations from debug-variant binaries, and a Tree-sitter oracle (2b) that parses pre-processed source into a syntactic function inventory with stable, content-based identifiers. Stage 3 joins the binary-side and source-side inventories using compiler-emitted line directives as a bridge, producing scored alignment pairs. Stage 4 decompiles the stripped binaries with Ghidra in headless mode, producing pseudocode and control-flow information. Stage 5 joins the oracle ground truth to the decompiler output by address overlap and applies the eight-gate quality funnel (Section 3.3) to assign each function a confidence tier. Stage 6 submits the resulting function records to LLMs for the naming task and scores predictions against ground truth.

## 3.2 Controlled Dataset Construction

The evaluation corpus is a researcher-authored micro-benchmark rather than a sample of real-world software. This is a deliberate trade-off: by controlling source authorship and build configuration, the pipeline isolates the effect of compiler optimization on alignment and naming difficulty without the problems (licensing, incomplete builds, mixed languages, missing ground truth) that affect corpus-scale datasets. The corpus comprises 15 synthetic C programs, ranging from single-file utilities to multi-file projects with shared headers, each designed to exercise compiler behaviours relevant to alignment difficulty such as cross-file calls, header sharing, struct passing, recursion, tail calls, and loop-invariant hoisting. The syntactic inventory contains 226 functions at the lowest optimization level. Each program is compiled across four optimization levels and three binary variants: a debug variant that emits DWARF for oracle extraction, a release variant, and a fully stripped variant that constitutes the realistic input presented to the decompiler and the LLM. This separates the binary used to derive ground truth from the binary the model actually sees.

## 3.3 The Eight-Gate Confidence Funnel

Our core methodological device is an explicit funnel that converts the fuzzy notion of reliable ground truth into a sequence of binary checks. Before the funnel, a function is screened for eligibility: it must possess a usable DWARF address range, must not be a non-target, must receive an ACCEPT verdict from the DWARF oracle, and must not be one of a frozen set of compiler- and linker-inserted infrastructure names. An eligible, successfully joined function must then pass all eight gates listed in Table 1 to be classified as high-confidence.

Table 1. The eight high-confidence gates, the condition each enforces, and the failure mode it is designed to exclude

| Gate | Condition for high confidence | Failure mode blocked |
|---|---|---|
| 1 | DWARF oracle verdict = ACCEPT | Unparseable or incomplete debug information |
| 2 | Alignment verdict = MATCH | Ambiguous or failed binary-to-source mapping |
| 3 | Single alignment candidate | Header replication producing multiple candidates |
| 4 | Overlap ratio ≥ 0.95 | DWARF range fragmentation under optimization |
| 5 | Strong address join (≥ 0.9) | Decompiler address-range mismatch |
| 6 | Not a noise function | Compiler-inserted infrastructure (thunks, proxies) |
| 7 | Control-flow graph not incomplete | Unresolved indirect jumps in control flow |
| 8 | No fatal decompiler warnings | Decompiler failure yielding unreliable pseudocode |

The gates are sequential and address distinct, stage-specific failure modes: oracle acceptance and alignment success are prerequisites for the later decompiler-fidelity and noise checks. Each gate records a diagnostic flag but never deletes the function record, preserving the full attrition audit trail.

---

produced byte-identical pseudocode for all 1,884 functions, as did a comparison against the original pipeline output produced months earlier on a different host.

Tier assignment combines eligibility, the strength of the address join, and the gate outcome into a single per-function label. GOLD functions pass all eight gates. SILVER functions are gold-eligible and achieve a strong address join but fail at least one high-confidence gate, with the failing gate recorded. In practice, the most common recorded causes for demotion from GOLD to SILVER are alignment overlap ratios between 0.70 and 0.94, multiple syntactic candidates, and decompiler-side noise classifications. BRONZE functions are joined, strongly or weakly, but are either not GOLD-eligible or achieve only a weak join. Functions that cannot be meaningfully joined receive no tier. All three tiers are evaluated independently, so that the experiment can be run at varying ground-truth strictness and the stability of conclusions across tiers examined directly. Tier membership correlates with structural simplicity as well as alignment quality (GOLD requires a single syntactic candidate, which itself selects for simpler functions), so tiers are best read as population strata for stratified reporting rather than as controlled treatment conditions.

## 3.4 Stable Cross-Optimization Keys and Paired Analysis

A function's identity must be tracked across optimization levels to ask whether the same function becomes harder to name as optimization increases. Binary-side identifiers are unsuitable for this. For instance, DWARF-derived function identifiers change for roughly 86% of functions between the two lowest optimization levels, so naive identifier-based tracking fails. Reforge instead anchors identity to the source declaration, using a key composed of the test case, declaration file, declaration line and column, and the normalized function name. The key degrades gracefully when a component is unavailable and never matches across optimization levels when the declaration site cannot be resolved. This stable key enables paired analysis: rather than comparing aggregate means across optimization levels (which compares different populations of functions), we compare the per-function score of the same function at the lowest and highest optimization levels, restricted to the cohort that remains evaluable at both. As Section 4 shows, this distinction matters in practice; paired and unpaired analyses yield qualitatively different conclusions about the effect of optimization.

## 3.5 Evaluation Metrics

Predictions are scored against ground-truth identifiers using two primary metrics. **Token F1** treats each name as a set of lowercase sub-tokens and computes the harmonic mean of token precision and recall; it is a lexical-overlap measure and therefore a lower bound on semantic recovery, since a semantically apt name that shares no tokens with the reference scores zero. **Exact match**, evaluated after normalization by lowercasing and removing separators, is a stricter all-or-nothing measure. To capture the analyst-assistance setting, in which a model proposes several candidates for human review, we additionally report a **top-K oracle** metric: given K ranked candidates, the score is that of the best candidate among them. The associated top-K uplift is the difference between the best-of-K and the top-1 score; it is non-negative by construction and, because the best candidate is selected post hoc, it represents an upper bound on the benefit an analyst or a downstream re-ranker could obtain from reviewing the shortlist.

# 4. PROOF-OF-CONCEPT EVALUATION

## 4.1 Experimental Setup

To demonstrate the substrate end to end, we evaluated seven contemporary LLMs spanning proprietary and open-weight families (Claude Sonnet 4.5, GPT-5.1, GPT-4o mini, DeepSeek R1, DeepSeek V3, Qwen3 Coder, and Llama-3.1 70B) on the single task of recovering a function's name from its decompiled, stripped form. Each model was run against each of the three confidence tiers at each of the four optimization levels, with models asked to return a ranked shortlist of three candidate names. The design spans 84 nominal model-tier-optimization cells, of which 77 are populated (the BRONZE tier contains no functions at -O0), together yielding 4,837 scored predictions. The decompiled input provided to the model included the raw pseudocode together with call relationships, a control-flow summary, and local-variable information.

## 4.2 Ground-Truth Attrition and Survivorship Bias

The first and most consequential finding concerns the substrate itself (RQ1). High-confidence yield, the fraction of gold-eligible functions that pass all eight gates, declines monotonically with optimization, from 87.2% at the lowest level to 74.2%, 68.7%, and 65.9% at successively higher levels (Figure 2). Two mechanisms drive the decline. At low optimization, the dominant losses are alignment-side: header replication produces multiple candidate matches that fail the uniqueness gate. As optimization increases, the failure mode shifts toward fidelity degradation, as inlining and code motion fragment the DWARF address ranges that anchor the mapping (Dramko et al., 2024). Independently, the evaluable function population shrinks from 220 functions at the lowest level to 132 at the highest, a 40% reduction attributable to inlining, folding, and dead-code elimination (Tan et al., 2025).

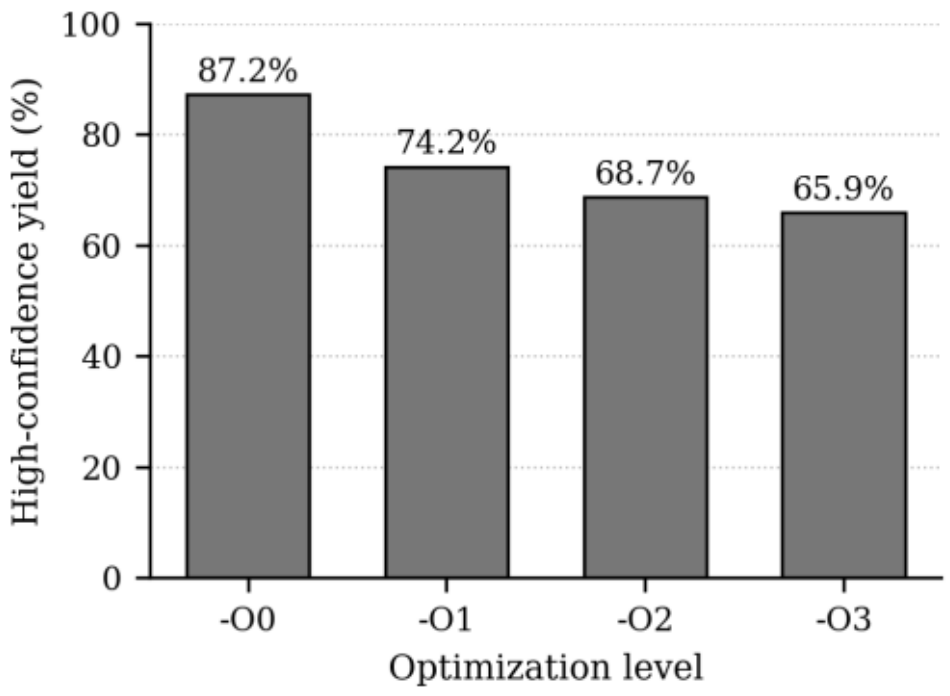

Figure 2. High-confidence yield by optimization level. The evaluable substrate contracts by more than twenty percentage points as optimization increases

This attrition is not merely a loss of sample size but confounds the most common comparison in the literature. Comparing aggregate mean accuracy across optimization levels compares different populations of functions, because those that survive to the highest level are a non-random, simpler subset of those present at the lowest. Using the stable key to construct a paired cohort (functions evaluable at both the lowest and highest levels) and comparing each function to itself, we find that mean per-function Token F1 changes only marginally, with per-model paired deltas ranging from −0.006 to +0.033. One-sided Wilcoxon signed-rank tests, corrected for multiple comparisons across the seven models, provide no robust evidence of systematic within-function decrease (all corrected $p \geq 0.209$). The apparent aggregate decay visible in unpaired means is therefore better explained by population shift than by individual functions becoming harder to name. The methodological implication is direct: paired analysis over a stable key should be the default when claiming optimization-level effects, and unpaired aggregate comparisons risk attributing a compositional artifact to a causal mechanism.

The tier transition matrices in Figure 3 corroborate this survivorship interpretation directly. The first optimization step is the most disruptive: 55 functions, a quarter of the -O0 population, leave the evaluable set entirely, while the later steps remove progressively fewer (24 and 10). Among the functions that remain evaluable, the large majority, between 75% and 96% depending on the transition, stay in their original confidence tier rather than being demoted. Attrition therefore acts mainly by removing functions from the population, not by degrading the ground-truth quality of those that survive, which is exactly the composition that produces survivorship bias in unpaired aggregate comparisons.

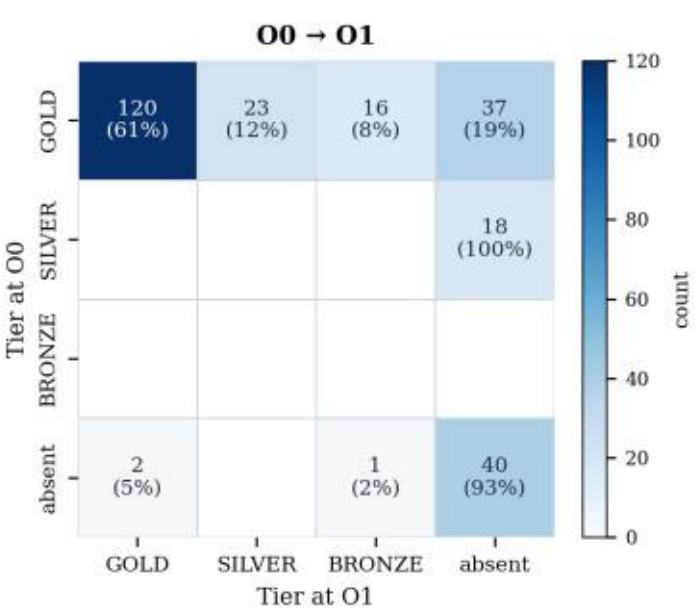

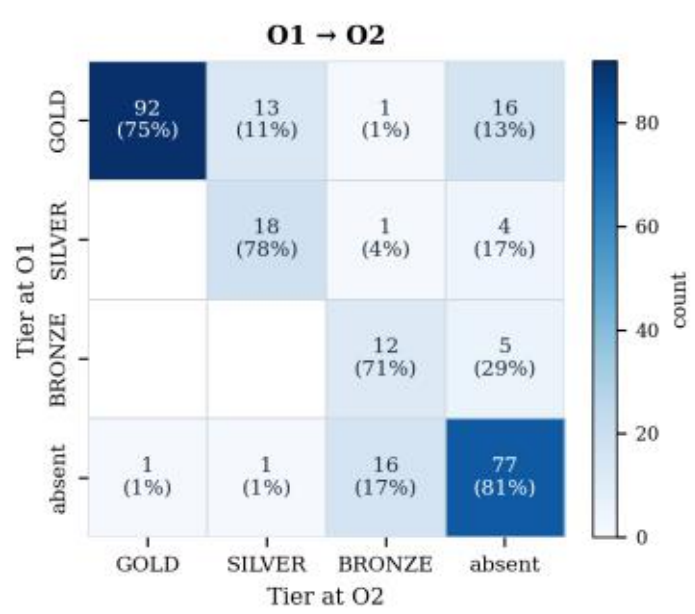

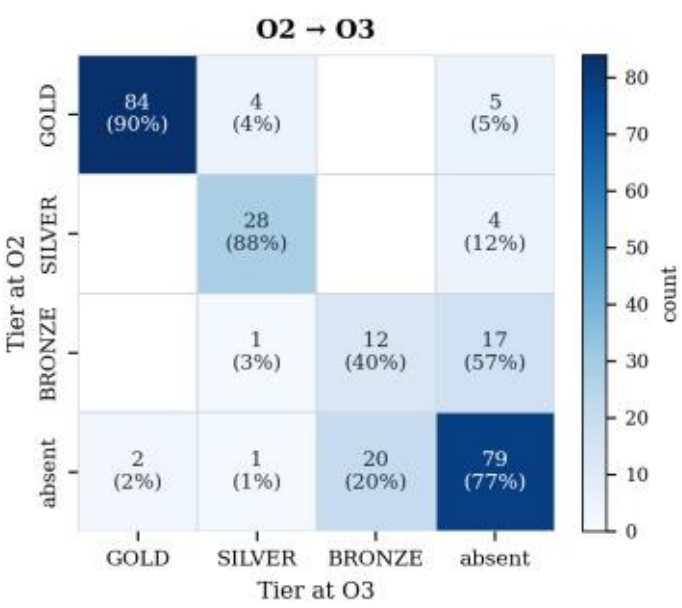

Figure 3. Tier transition matrices across consecutive optimization levels (O0 to O1, O1 to O2, O2 to O3). Each cell counts functions moving from a row tier at the lower level to a column tier at the higher level; the absent row and column denote exit from or entry to the evaluable set. The heavy diagonal mass shows that most surviving functions retain their tier, while attrition is concentrated in the absent column

## 4.3 Function-Naming Performance

On the naming task itself (RQ2), performance is modest and consistent with prior reports of weak fine-grained identifier recovery (Jin et al., 2023). Mean Token F1 across the seven models lies in a narrow band of roughly 0.18 to 0.29, with a median of zero: on more than half of functions the top prediction shares no tokens with the reference. Exact-match rates are correspondingly low, between 1.9% and 7.1% across models, so only about one prediction in fifteen to fifty reproduces the original name exactly after case and separator normalization. The narrowness of the band across very different model families, together with moderate-to-strong cross-model agreement on which functions are easy or hard (Kendall's W = 0.66 for functions whose confidence tier is stable across optimization levels and 0.77 for tier-hopping functions), suggests that naming difficulty here is driven more by the information content of the decompiled input than by model-specific capability. We report these figures to characterize what the substrate measures, not to rank the models; the small synthetic corpus and lexical metric preclude capability conclusions.

To probe how much semantic recovery the lexical metric misses, we re-scored all 4,837 stored predictions with an embedding-based similarity and compared each score against a random-pair baseline[2]. A code-specific encoder, CodeBERT, separated matched from mismatched name pairs only weakly, reaching an area under the ROC curve (AUC) of 0.64, or 0.66 after whitening away its strong anisotropy, so we instead report a sentence-encoder metric, bge-small-en-v1.5. That metric reaches an AUC of 0.78, against a random-pair baseline whose mean cosine is 0.53 and whose 95th percentile is 0.65. It is most informative on the predictions that Token F1 scores as outright failures. Among predictions with a Token F1 of zero, 19.3% exceed the baseline 95th percentile, where only 5% would be expected by chance. Net of that chance rate, roughly one in seven lexically-zero predictions is in fact semantically apt. The clearest cases are near synonyms: the model predicts "matrix_multiplication" for a function named "mat_multiply", which shares no tokens with the reference yet names the same operation. This confirms that Token F1 acts as a lower bound on semantic recovery, and it bounds the resulting underestimate at ~8% of all predictions. The remaining zero-scoring predictions are genuinely unrelated, sitting on the random baseline as a group, with a mean cosine of 0.54 against the baseline of 0.53. Finally, the two metrics agree on the relative standing of the models, at a Spearman rank correlation of 0.86 between per-model mean Token F1 and per-model mean semantic score. So neither the absolute performance figures nor our decision not to rank models on this corpus changes under the semantic metric.

## 3.5 Tier Strata Revisited

Stratified by tier, GOLD-tier Token F1 means are slightly higher (roughly 0.20 to 0.32) than the all-tier band, and this tracks a real structural difference between the populations (Figure 4). Among the two strong-join tiers, GOLD functions are consistently simpler than SILVER, in both mean and median and on every measure, consistent with the single-candidate gate that defines GOLD; BRONZE is a separate, bimodal case, split between small weakly-joined stubs (a majority are single-block, branchless fragments of only a few lines, which give it the lowest medians) and a tail of larger functions (which lift its mean above GOLD's). Tier complexity therefore does not track the ground-truth-quality ordering, so the tier-level performance gap is best read as a population difference rather than a causal effect of ground-truth quality.

[2] Reproduction scripts, per-prediction semantic scores, and the decompiler determinism evidence are available at https://github.com/NicolasKol/reforge/tree/master/data/results/paper_camera_ready.

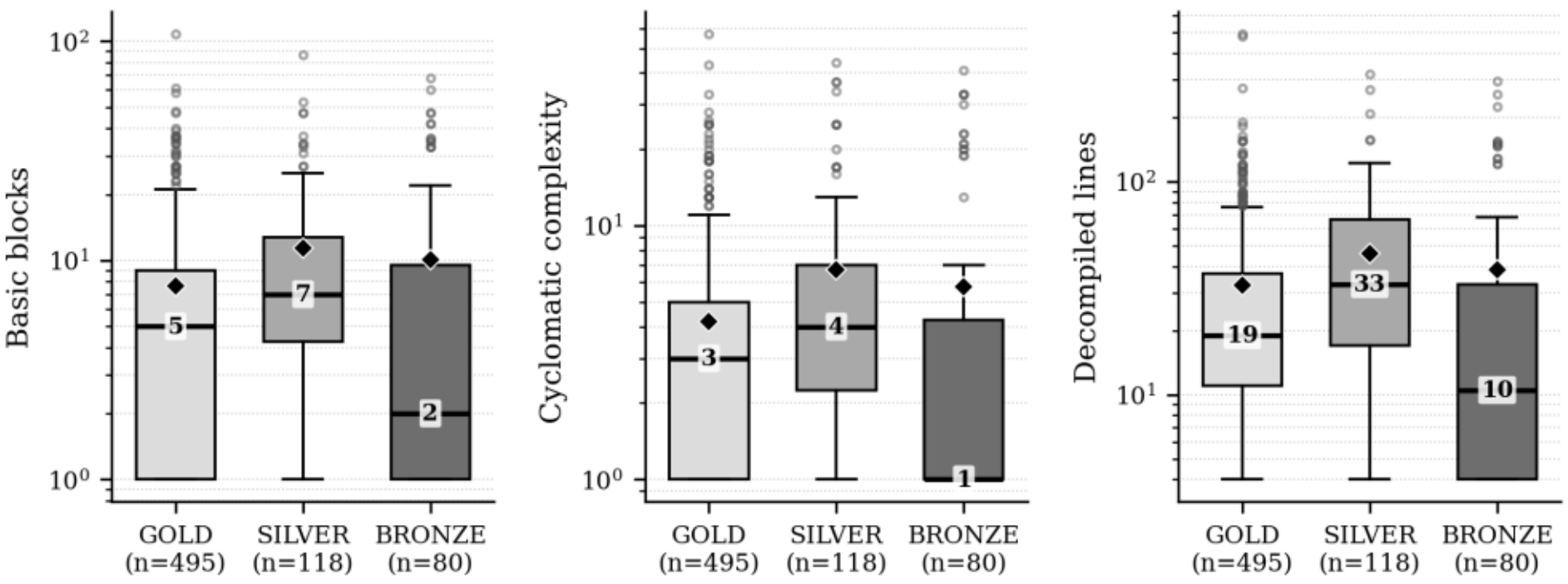

Figure 4. Structural profile of the three confidence tiers, pooled over optimization levels (log scale; n per tier shown). Each box spans the interquartile range; the central line is the median (value labelled), the diamond is the mean, whiskers reach 1.5 times the IQR, and open circles mark individual outlier functions beyond that. GOLD is consistently simpler than SILVER; BRONZE has the lowest medians but is bimodal, mixing small weak-join stubs (which give it those low medians) with a tail of larger functions (which lift its mean above GOLD's).

## 4.4 Shortlist Utility for Analyst Workflows

The third dimension (RQ3) examines whether models are more useful when they propose several candidates rather than one. Across models, moving from the single top-ranked prediction to the best of three candidates yields a positive Token F1 uplift for between 13.6% and 26.7% of functions (Figure 5), with mean uplift magnitudes of roughly 0.06 to 0.12. Because the best candidate is selected by an oracle, this is an upper bound on achievable benefit. Nonetheless, the pattern is informative. In a substantial fraction of improved cases the best candidate appears at rank two or three rather than rank one, indicating a ranking or calibration gap: the models can generate a closer name than they place first.

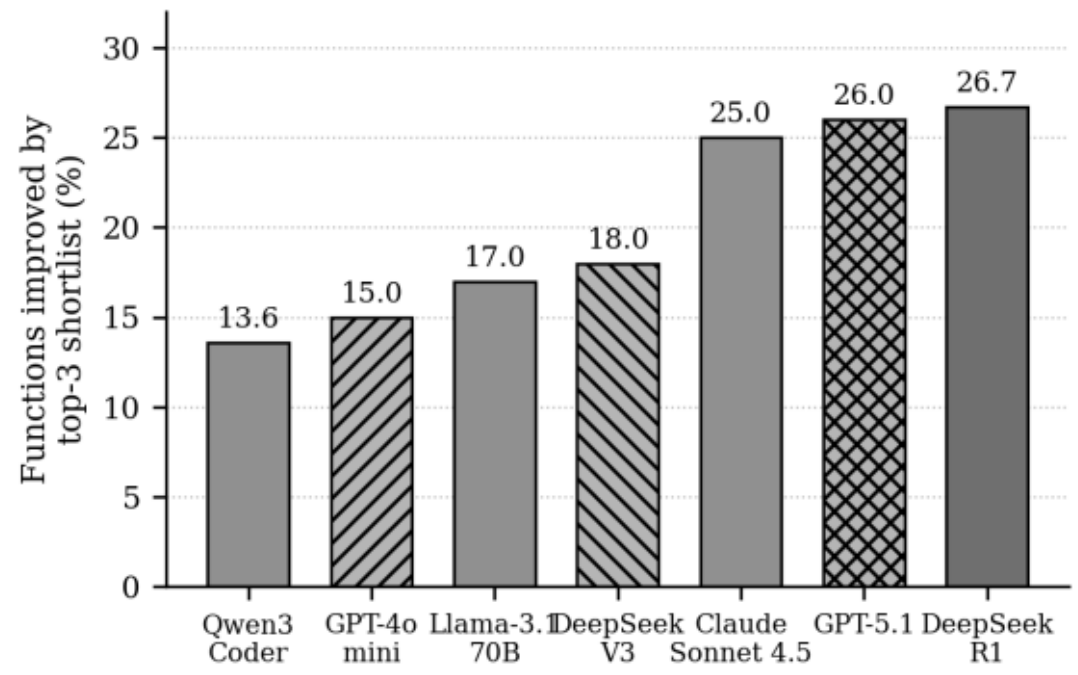

Figure 5. Fraction of functions whose Token F1 improves with the best-of-three approach vs. top-ranked prediction only

The oracle uplift is an upper bound, so we also estimated a *lower* bound on what a practical re-ranker can achieve. To do this, we replayed all 4,837 stored predictions through two selection strategies that must choose one of the three candidates without seeing the ground truth. The first strategy selects the candidate with the highest model-reported confidence. It reproduces the top-ranked prediction in every single case, because no model ever orders its confidence scores differently from its ranks, and it therefore captures none of the oracle uplift. The second strategy matches candidates against the function body: it selects the candidate whose name tokens share the most characters with identifiers harvested from the function's decompiled pseudocode. This heuristic performs slightly worse than simply keeping the top-ranked prediction. Pooled over all predictions, mean Token F1 drops by 0.012; the heuristic improves 3.3% of functions but worsens 4.9%, and where it overrides the top rank it is wrong more often than right. The cause is instructive. Once Ghidra's automatically generated names are excluded, the body of a stripped function retains a median of only two meaningful identifier tokens, so in 69% of cases the heuristic cannot separate the candidates at all, and the cases it can separate are dominated by superficial matches. For comparison, choosing among the three candidates at random lowers mean Token F1 by 0.073, which confirms that the models' own ranking is already strongly informative. Taken together, the upper and lower bounds bracket the practically achievable shortlist benefit. That is, intra-function heuristics recover essentially none of the oracle uplift, and closing the gap will require signals from outside the function body. This resonates with the operational reliability concern raised by recent threat reporting (Anthropic, 2025): where a model's single best guess is unreliable, surfacing a small ranked shortlist for human adjudication is better matched to how analysts actually work, and integrating such shortlists into decompiler tooling is a low-cost design change that hedges against single-answer error.

## 5. DISCUSSION, CONCLUSION AND FUTURE WORK

This paper argued that the principal obstacle to fair benchmarking of LLM reverse-engineering capability is the construction and auditing of the evaluation basis, not model capability itself. With Reforge, a provenance-tracked pipeline with a deterministic ground-truth core built on the function-level, we make alignment yield and attrition auditable, and the eight-gate confidence funnel operationalizes alignment uncertainty as explicit, stratified output. Using the method, we showed that high-confidence yield falls from 87.2% to 65.9% across optimization levels and that unpaired optimization comparisons overstate performance decay through survivorship bias, a confound removed by paired analysis over a stable, source-anchored key. A proof-of-concept evaluation of seven LLMs on function naming demonstrated the method end to end.

The findings divide cleanly into two kinds, and the distinction matters for how they should be used. The methodological results (optimization-driven attrition, its shifting failure modes, and the survivorship-bias confound) are properties of the compilation-and-alignment process, not of the particular corpus or models. They are expected to hold for any toolchain that emits DWARF-equivalent metadata, and they carry a concrete prescription: benchmarks should report alignment yield and per-stage attrition alongside accuracy, and should use paired analysis over a source-anchored key whenever they claim optimization effects. The empirical performance results (the absolute Token F1 values and the model ordering) do not generalize beyond the present setting and are offered only as evidence that the baseline functions.

Read this way, we reframe where effort should be spent. The limiting factor on the validity of a function-level benchmark is not, in the first instance, the model under test but the reliability of the mapping that defines a correct answer. Making that reliability explicit changes what a benchmark reports: not a single accuracy number per optimization level, but a yield-stratified, provenance-tracked account in which ground-truth uncertainty is visible to the user. For tool builders, the shortlist result suggests a practical near-term design: presenting a small ranked set of name candidates within a decompiler, rather than a single suggestion, better matches the analyst's hypothesis-testing workflow and hedges against the unreliability that makes single best guesses risky in operation.

The empirical scope is the main limitation which defines the work ahead. The corpus is small, synthetic, and built with a single compiler for a single architecture and decompiler; the absolute performance figures should not be generalized to production binaries, other toolchains, or other languages, and Token F1 captures only lexical overlap. The confidence tiers conflate alignment quality with function complexity, and the reported shortlist benefit is an oracle upper bound. Each limitation is a concrete next step. The most immediate is scaling the pipeline to real-world open-source corpora compiled via reproducible build recipes such as those provided by the Assemblage infrastructure (Liu et al., 2024), to test whether the qualitative conclusions hold at larger scale and greater structural diversity. Cross-toolchain replication (Clang/LLVM, architectures such as ARM64, and additional decompilers) would establish the generality of the attrition findings, a gate-threshold sensitivity analysis would quantify the yield–noise trade-off the funnel encodes, and closing the gap between the oracle bound and the near-zero benefit of intra-function heuristics shown in Section 4.4 calls for a re-ranker that draws on signals from outside the function body, such as call-graph context or cross-reference frequency (Xu et al., 2023). We intend the pipeline to serve as a reusable, openly available artifact supporting these extensions.

**Data and Code Availability.** The Reforge pipeline, the synthetic corpus, and the reproducible evaluation notebooks are openly available at https://github.com/NicolasKol/reforge.

**AI Use Disclosure.** This paper condenses original research of the first author's thesis. AI tools were used throughout the production of this thesis, spanning pipeline development, data analysis, literature navigation, and text drafting. The present manuscript faithfully reflects that research; generative AI tools assisted only with editorial adaptation and language during its initial preparation and were not used to generate substantive new content, results, or claims. AI assistance in the revision of the paper extended to data analysis, figure generation, script authoring, and drafting**.** All results were independently human-reproduced and generated texts are human-checked.

**Acknowledgements.** The authors thank three anonymous reviewers for their substantive remarks which helped to significantly augment and improve the paper in its revised version.